\def\ni{\noindent}

\def\etal{{\it et~al.~}}
\def\bsax{{\it BeppoSAX~}}
\def\ginga{{\it Ginga~}}

\def\rxte{{\it RXTE~}}

\def\heao1{{\it HEAO-1~}}

\def\cts{~{\rm cts~s}^{-1}~}
\def\h0{~{\rm H_0 = 50~km~s}^{-1}\ {\rm Mpc}^{-1}~h_{50}~}

\def\cts{~{\rm counts~s}$^{-1}$~}

\documentclass[12pt,preprint]{aastex}
\begin{document}

\newcommand{\lessim}{\ \raise -2.truept\hbox{\rlap{\hbox{$\sim$}}\raise5.truept
    \hbox{$<$}\ }}

\title{Hard X-Ray Excess in the Coma Cluster: a Reply to Rossetti \& Molendi
(astro-ph/0702417) }

\author{Roberto Fusco-Femiano$^a$, Raffaella Landi$^b$, Mauro Orlandini$^b$
\footnote{$^a$Istituto di Astrofisica Spaziale e Fisica Cosmica
(IASF/Roma), INAF, via del Fosso del Cavaliere, I--00133 Roma,
Italy - roberto.fuscofemiano@iasf-roma.inaf.it; $^b$IASF/Bologna,
INAF, via Gobetti 101, I--40129 Bologna, Italy -
landi@iasfbo.inaf.it - orlandini@iasfbo.inaf.it}}

%\centerline{\large{\bf {Hard X-Ray Excess in the Coma Cluster:}}}
%\centerline{\large{\bf {a Reply to Rossetti \& Molendi
%(astro-ph/0702417)}}}

%\medskip
%---- ENTRY 2 ----------------------------------------------------------
%
% Type below the Principal Investigator (PI) initial(s) and family name
%\centerline{PI: {\textit{R.Fusco-Femiano}}}

%\ni \centerline{\textit{Roberto Fusco-Femiano$^1$, Raffaella
%Landi$^2$ \& M.Orlandini$^2$}}

%\bigskip
%\centerline{$^1$IASF-Roma/INAF), Roma, Italy
%(e-mail:roberto.fuscofemiano@iasf-roma.inaf.it)}

%\centerline{ $^2$IASF-Bologna/INAF, Bologna, Italy}

%\bigskip

%\noindent {\bf 1. Abstract}

We read with surprise the comment "The Coma Cluster hard X-ray
spectrum revisited: still no evidence for a hard tail" by Rossetti
\& Molendi (hereafter RM07) that appeared on astro-ph
(astro-ph/0702417), in answer to our paper Fusco-Femiano, Landi \&
Orlandini 2007 (hereafter FF07). Indeed the points raised by these
authors are extensively explained in the Fusco-Femiano \etal 2004
(hereafter FF04), FF07 and also Fusco-Femiano, Landi \& Orlandini
2005 (hereafter FF05) that regards the PDS data analysis of Abell
2256. A more careful reading of these papers would have avoided to
Rossetti \& Molendi to write the comment to our paper FF07. We
want to stress our surprise because we do feel that the ArXig.org
server should not be a forum for discussion but a server that
collects preprints submitted (or, even better, already accepted)
to scientific journals. Sending comments like RM07 that cast
doubts on papers that passed a serious refereeing process has the
only effect to increase the "noise" at detriment of the Science.
Nevertheless, we feel forced to reply to this preprint, but it is
of course our last reply because we do not want to bore the
community any longer with this dispute. We limit to report here a
short discussion regarding the conclusions of RM07. More details
can be found in the above cited papers.

\medskip
\par\noindent
In FF07 we have shown that also using the same software package
(SAXDAS) used by Rossetti \& Molendi 2004 (hereafter RM04) in
their PDS data analysis of the Coma cluster, it is possible to
obtain the same results obtained in FF04 using a different package
(XAS). In particular, we have confirmed the presence of a
nonthermal excess with respect to the thermal emission by the
intracluster gas at about the same confidence level ($\sim
4.8\sigma$ with XAS and $\sim 4.6\sigma$ with SAXDAS).
\medskip
\par\noindent
The conclusions reported in RM07 regard:
\par\noindent
a) "\textit{the choice of the background, since two OFF fields are
available}". We have already explained in FF04 and FF07 that we
consider only the -OFF sky position for the determination of the
background for the presence in the +OFF position of the variable
source BL Lac 1ES 1255+244. In the two \bsax observations of the
Coma cluster (OBS1 and OBS2) the comparison between the two
accumulated backgrounds (difference between the +OFF and --OFF
count rate spectra) showed that for OBS1 the difference was
compatible with zero while for the longer, more sensitive OBS2,
there was an excess of $0.064\pm 0.021$ \cts that is consistent
with our analysis of the BL Lac observed by Beckmann \etal 2002.
Moreover, this difference between the count rates of +OFF and -OFF
background persists for all the whole OBS2 of about 300 ks. This
justifies our decision to exclude the +OFF position for the
background determination. However, we have computed in FF04 the
c.l. of the excess considering the two sky directions. We obtain
$\sim 3.9\sigma$ that confirms the presence of a hard excess in
the Coma cluster spectrum.
\medskip
\par\noindent
b) "\textit{the choice of the temperature value for the thermal
value}". It is necessary to consider the average temperature in a
region of possibly the same size of the field of view of the PDS
detector (FWHM = $1.3^{\circ}$). We have considered the gas
temperature measured by \ginga (8.11$\pm$0.07 keV, 90\%; David
\etal 1993) that has a FOV comparable to that of the PDS. The
\ginga value has been confirmed by \rxte that reports 7.9$\pm$0.03
keV (Rephaeli \& Gruber 2002) with a FOV of $\sim 1^{\circ}$. The
choice to consider the average temperature in a region of size
comparable to that of the PDS is supported also by theoretical
investigations (e.g., Brunetti \etal 2001; Colafrancesco,
Marchegiani \& Perola 2005). It is expected that the nonthermal
emission is mainly contributed by the outer regions (30-50 arcmin)
of the cluster volume which contain the large majority of the
relativistic electrons. However, in FF04 we have shown that for a
temperature of 8.25 keV (RM04 uses 8.21$\pm$0.16 keV) the excess
is at the level of $\sim 4.6\sigma$. To satisfy Rossetti \&
Molendi we have computed the excess for 8.4 keV, the upper limit
of their reported interval in RM07. We obtain a c.l. for the
excess of $\sim 4.15\sigma$ (observed count rate =
0.3489$\pm$0.0153 \cts, model predicted rate = 0.2854 \cts) that
confirms the presence of a hard tail in the Coma cluster.
\medskip
\par\noindent
c) "\textit{the choice of the observation, the two spectra
separately or their sum}". We outline that the combined spectrum
is obtained taking into account the different exposure times of
OBS1 and OBS2 (technically speaking, the sum of spectra must be
always performed in \textit{counts} units and not in \textit{rate}
units). As a consequence, OBS2 has a predominant importance also
in the combined spectrum. Besides, we outline also here (see FF07)
the different behaviour of the two packages after any South
Atlantic Geomagnetic Anomaly (SAGA) passage. The SAXDAS package
removes 5 minutes after any passage, while XAS eliminates the time
necessary to allow the PDS high voltage to reach the correct
levels after its shut-down during the SAGA passage. So, some
spurious events could be present in the SAXDAS analysis, in
particular for the longer OBS2 observation. The XAS package
results to be more conservative with an exposure time lower than
about 2 ksec with respect to SAXDAS for the different time removal
after the SAGA passage.

\medskip
\par\noindent
Finally, besides the 3 points discussed above, we would like to
remark briefly the following points:

\textit{i)} A rigorous selection of the events it is crucial in
order to eliminate the presence of any spikes able to introduce
noise that hides the presence of a nonthermal excess with respect
to the thermal radiation. We have a significant increase of the
c.l. of the excess (from $\sim 2.9\sigma$ to $\sim 4.2\sigma$)
when we consider in the SAXDAS analysis (FF07) the same time
windows used in the XAS analysis (FF04). The selection of the
events are obtained in the XAS analysis with an automatic
selection followed by a visual check in order to eliminate all the
remaining spikes. We are sure that also Rossetti \& Molendi would
have confirmed the presence of a nonthermal excess considering the
same time windows used in FF04.

\textit{ii)} RM07 report that we do not provide in FF07 details of
the analysis regarding the systematic errors and that also
Nevalainen \etal 2004 find that there is a systematic difference
between the OFF fields. We stress here that the systematic errors
are discussed in detail in FF05, and then in FF07, taking into
account the \textit{whole} sample of the PDS observations and that
the referee of FF05 was Dr. J.Nevalainen. In particular, the
referee was in agreement with our analysis on the whole sample of
PDS pointings (868, while RM04 consider only 69 observations)
regarding the possible systematic difference between the OFF
fields reported in RM04 and RM07. Our analysis gives a value of
$(5.3\pm 6.3)\times 10^{-3}$\cts, consistent with no contamination
at all.

\textit{iii)} RM07 continue to report (we hope for the last time
!!) the \textit{trivial} error committed in Fusco-Femiano \etal
1999 that has been widely corrected in the subsequent analysis
(FF04).

\bigskip
\textbf{References}
%\medskip

\ni Beckmann, V. \etal 2002, A\&A, 383, 410

\ni Brunetti, G., Setti, G., Feretti, L., \& Giovannini, G. 2001,
MNRAS, 320, 365

\ni Colafrancesco, S., Marchegiani, P., \& Perola, G.C. 2005,
A\&A, 443, 1

\ni David, L.P., Slyz, A., Jones, C., Forman, W., \& Vrtilek, S.D.
1993, ApJ, 412, 479

\ni Fusco-Femiano, R., Dal Fiume, D., Feretti, L. Giovannini, G.,
Grandi, P., Matt, G., Molendi, S., \& Santangelo, A. 1999, ApJ,
513, L21

\ni Fusco-Femiano, R. Orlandini, M., Brunetti, G., Feretti, L.,
Giovannini, G., Grandi, P. \& Setti, G. 2004, ApJ, 602, L73 (FF04)

\ni Fusco-Femiano, R., Landi, R. \& Orlandini, M. 2005, ApJ, 624,
L69 (FF05)

\ni Fusco-Femiano, R., Landi, R. \& Orlandini Mauro 2007, ApJ,
654, L9 (FF07)

\ni Nevalainen, J., Oosterbroek, T., Bonamente, M., \&
Colafrancesco, S. 2004, ApJ, 608, 166

\ni Rephaeli, Y. \& Gruber, D.E. 2002, ApJ, 579, 587

\ni Rossetti, M. \& Molendi, S. 2004, A\&A, 414, L41 (RM04)

\ni Rossetti, M. \& Molendi, S. 2007, astro-ph/0702417 (RM07)

\ni

%---- ENTRY 4 ----------------------------------------------------------
%
% This is the place where the proposed program is to be described.
% This description is composed of two different sections.
%
% A) Scientific rationale: scientific background of the project,
% pertinent references; previous work plus justification for present
% proposal.
%
% B) Immediate objective of the proposal: state what is actually going
% to be observed and what shall be extracted from the observations, so
% that the feasibility becomes clear.
%

%\ms \noindent {\bf 2. Description of the proposed program\\}
%\noindent {\sl A) Scientific Rationale:}

\end{document}